\documentclass[%
reprint,
superscriptaddress,
showpacs,
showkeys,
amsmath,amssymb,
aps,
prb,
]{revtex4-1}
\usepackage{graphicx,amssymb,amsmath,epsfig}
\usepackage{color}
\usepackage{comment}
\usepackage[usenames,dvipsnames,svgnames,table]{xcolor}
\usepackage[lofdepth,lotdepth,caption=false]{subfig}
\usepackage[pdfstartview=FitH]{hyperref}
\usepackage{hyperref}
\newsavebox{\bigleftbox}
\hypersetup{
 colorlinks=true,
 citecolor=Blue,
 linkcolor=Blue,
 urlcolor=Blue
}

\begin{document}

\title{On the Mechanical, Electronic, and Optical Properties of 8-16-4 Graphyne: A 2D Carbon Allotrope with Dirac Cones}
\author{Raphael M. Tromer}
\affiliation{Applied Physics Department, State University of Campinas, Campinas-SP, 13083-970, Brazil}
\affiliation{Center for Computing in Engineering \& Sciences, Unicamp, Campinas-SP, Brazil}
\author{Marcelo L. Pereira J\'unior}
\affiliation{University of Bras\'{i}lia, Faculty of Technology, Department of Electrical Engineering, Bras\'{i}lia, Brazil.}
\author{Kleuton A. L. Lima}
\affiliation{Department of Physics, Federal University of Rio Grande do Norte, Natal, Rio Grande do Norte, Brazil.}
\author{Alexandre F. Fonseca}
\affiliation{Applied Physics Department, State University of Campinas, Campinas-SP, 13083-970, Brazil}
\affiliation{Center for Computing in Engineering \& Sciences, Unicamp, Campinas-SP, Brazil}
\author{Luciano R. da Silva}
\affiliation{Department of Physics, Federal University of Rio Grande do Norte, Natal, Rio Grande do Norte, Brazil.}
\author{Douglas S. Galv\~ao}
\affiliation{Applied Physics Department, State University of Campinas, Campinas-SP, 13083-970, Brazil}
\affiliation{Center for Computing in Engineering \& Sciences, Unicamp, Campinas-SP, Brazil}
\author{Luiz. A. Ribeiro Junior}
\email{ribeirojr@unb.br}
\affiliation{University of Bras\'ilia, Institute of Physics, Bras\'ilia-DF, 70910-970, Brazil}

\date{\today}

\begin{abstract}
Due to the success achieved by graphene, several 2D carbon-based allotropes were theoretically predicted and experimentally synthesized. We used density functional theory and reactive molecular dynamics simulations to investigate the mechanical, structural, electronic, and optical properties of 8-16-4 Graphyne. The results showed that this material exhibits good dynamical and thermal stabilities. Its formation energy and elastic moduli are -8.57 eV/atom and 262.37 GPa, respectively. This graphyne analogue is a semi-metal and presents two Dirac cones in its band structure. Moreover, it is transparent, and its intense optical activity is limited to the infrared region. Remarkably, the band structure of 8-16-4 Graphyne remains practically unchanged at even moderate strain regimes. As far as we know, this is the first 2D carbon allotrope to exhibit this behavior.
\end{abstract}



\maketitle

\section{Introduction}

Following the discovery of graphene in 2004 \cite{novoselov2004electric}, there is a renewed interest in 2D carbon materials \cite{Ando2009,Sangwan2018,Song2018,kumar2018recent}. Their physicochemical properties can be controllable depending on the synthesis process \cite{lu2013two,wang2016electronic}. Several 2D carbon allotropes have been proposed \cite{enyashin2011graphene,wang2015phagraphene,wang2018popgraphene,zhuo2020me,karaush2014dft,zhang2019art,PhysRevB.70.085417,Alsayoud2018,tromer2021thiophene,diboron,jana2021emerging}, and a few of them have been already experimentally realized \cite{toh2020synthesis,fan2021biphenylene,doi:10.1021/jacs.2c06583,hou2022synthesis}. Among the recently synthesized structures, it is worth mentioning monolayer amorphous carbon (MAC) \cite{toh2020synthesis}, 2D biphenylene network (BPN) \cite{fan2021biphenylene}, monolayer fullerene network (2DC$_{60}$) \cite{hou2022synthesis}, and the multilayer $\gamma$-Graphyne \cite{doi:10.1021/jacs.2c06583}. 

MAC and BPN share some graphene properties, such as a zero semi-metal bandgap and the Dirac cones (corresponding to a linear dispersion). MAC has randomly distributed defects with five, six, seven, and eight rings of carbon atoms. Its lattice arrangement differs from disordered graphene. The BPN lattice contains a periodic arrangement of four, six, and eight carbon rings. On the other hand, $\gamma$-Graphyne and 2DC$_{60}$ present a small direct bandgap of 0.48 eV \cite{doi:10.1021/jacs.2c06583} and direct semiconducting bandgap of 1.6 eV \cite{hou2022synthesis}, respectively. 

The lattice structure of $\gamma$-Graphyne can be understood as graphene uniformly expanded by inserting two-carbon acetylenic units between all the aromatic rings. The graphyne structures were predicted in 1987 \cite{Baughman1987}. Graphynes are a generic name for structures where acetylene groups are inserted into single bonds. Graphyne-based molecules (fullereneynes) \cite{coluci2003families}, nanotubes \cite{baughman1993crystalline}, and nanoscrolls \cite{solis2018structural} have been reported in the literature.

2DC$_{60}$ crystals are formed by C$_{60}$ polymers covalently bonded in a planar configuration. This clustering mechanism yielded two stable crystals of polymeric C$_{60}$ in closely packed quasi-hexagonal and quasi-tetragonal phases \cite{hou2022synthesis}. $\gamma$-Graphyne and 2DC$_{60}$ overcome the problem of a null bandgap shown by other 2D carbon-based materials, which limits their applications in digital electronics.

Other carbon-based materials (and materials composed of different atomic species) with Dirac cones were proposed \cite{Liu2022,Zhang2021_C,wang2015phagraphene,Adjizian2014,Chen2020}. S-graphene \cite{Bandyopadhyay2020}, and 6,6,12-graphyne \cite{Wang2014_C} are examples of 2D carbon materials that can present an additional Dirac point in their band structure profiles. An external strain applied to these materials can tune the Dirac point. In moderate strain regimes, the two Dirac cones merge into only one cone. Under high strain rates, it was observed a bandgap transition from semimetallic to semiconductor, and the Dirac cones disappeared with the opening of the bandgap. 

It might be of interest for some applications in flexible electronics that the optoelectronic properties of the material do not undergo any substantial change when subjected to external stress. To our knowledge, no discussions on a 2D carbon-based material with this property have been presented in the literature. Therefore, this work aims to fill this gap.

Herein, we study the mechanical, structural, electronic, and optical properties of 8-16-4 Graphyne \cite{bandyopadhyay20218}, a 2D carbon-based material that presents two Dirac cones in its band structure. For simplicity, from now on we named it Sun-Graphyne (S-GY) (see Figure \ref{fig1}), considering that its lattice structure resembles other graphyne varieties and contains two rings with eight atoms  \cite{Baughman1987,Kang2019,Desyatkin2022}. Moreover, this name is also due to the atomic arrangement of its unit cell, which resembles the sun drawing images. We investigated the S-GY electronic, mechanical, and optical properties using density functional theory (DFT) and \textit{ab initio} and classical molecular dynamics (MD) methods. Our analyses revealed that S-GY is thermally stable up to high temperatures. 

The S-GY structure composes the class of theoretically designed graphyne structures. We started from octa-graphene \cite{sheng2012octagraphene}, and acetylene groups are inserted into the bonds of the square octa-graphene rings. It differs from the so-called T-graphyne (also called octa-graphyne) \cite{farooq2022topological_octa,jana2019acetylenic_T}, which contains only one eight-atom ring. S-GY differs from other Dirac materials because its electronic band structure does not significantly change under a moderate strain regime. 

\section{Methodology}

\subsection{DFT Calculations}

DFT simulations were used to study the S-GY electronic and optical properties. The simulations were carried out using the SIESTA code \cite{Soler2002}, and within the framework of the generalized gradient approximation (GGA) with the Perdew-Burke-Ernzenhof (PBE) exchange-correlation functional \cite{Perdew1996,ernzerhof1999assessment}. 

Norm-conserving Troullier-Martins pseudopotentials were used to describe the core electrons \cite{Troullier1991}. The calculations considered van der Waals (vdw-DFT) corrections \cite{vdw1,vdw2,vdw3}. Double-zeta plus polarization (DZP) was used as the basis set. A vacuum region of 20 \r{A} was employed to prevent spurious interactions among the periodic images. We assumed a cut-off value of 300 eV for the kinetic energy. The k-grid was $10\times10\times1$ for geometry optimizations and $30\times30\times1$ for electronic and optical calculations, respectively.

We used \textit{ab initio} MD (AIMD) approach in simulations considering finite temperatures, with an NVT ensemble with an integration time step of 1.0 fs, for a total simulation time of 2 ps. A Nosé-thermostat maintains the temperature constant when equilibrium is reached \cite{Evans1985}.

To perform the optical calculations, we considered a standard external electric field of 1.0 V/\r{A} along the x, y, and z directions \cite{Tromer2017}. From the Kramers-Kronig relation and Fermi's golden rule, we can obtain the real $\epsilon_1$ and imaginary parts of the dielectric constant: 
$\epsilon_2$, respectively:
\begin{equation}
\epsilon_1(\omega)=1+\frac{1}{\pi}P\displaystyle\int_{0}^{\infty}d\omega'\frac{\omega'\epsilon_2(\omega')}{\omega'^2-\omega^2},
\end{equation}
where $P$ is the Cauchy principal value, and
\begin{equation}
\epsilon_2(\omega)=\frac{4\pi^2}{V_\Omega\omega^2}\displaystyle\sum_{i\in \mathrm{VB}, \, j\in \mathrm{CB}}\displaystyle\sum_{k} W_k \, |\rho_{ij}|^2 \, \delta	(\varepsilon_{kj}-\varepsilon_{ki}- \hbar \omega),
\end{equation}
where $W_k$ is the k-point weight in the reciprocal space, $\rho_{ij}$ the dipole transition matrix element, c$\omega$ the photon frequency, and $V_\Omega$ the unit cell volume. VB and CB are the valence and conduction bands, respectively \cite{tromer2021thiophene}. 

Once the real and imaginary parts of the dielectric constant are obtained, the other relevant optical coefficients, such as the absorption coefficient $\alpha$, the refractive index ($\eta$), and reflectivity ($R$), can be derived as:  
\begin{equation}
\alpha (\omega )=\sqrt{2}\omega\bigg[(\epsilon_1^2(\omega)+\epsilon_2^2(\omega))^{1/2}-\epsilon_1(\omega)\bigg ]^{1/2},
\end{equation}
\begin{equation}
\eta(\omega)= \frac{1}{\sqrt{2}} \bigg [(\epsilon_1^2(\omega)+\epsilon_2^2(\omega))^{1/2}+\epsilon_1(\omega)\bigg ]^{2},
\end{equation}
and
\begin{equation}
R(\omega)=\bigg [\frac{(\epsilon_1(\omega)+i\epsilon_2(\omega))^{1/2}-1}{(\epsilon_1(\omega)+i\epsilon_2(\omega))^{1/2}+1}\bigg ]^2.
\end{equation}

\subsection{Classical MD Simulations}

\begin{figure*}[!htb]
\centering
\includegraphics[width=0.7\linewidth]{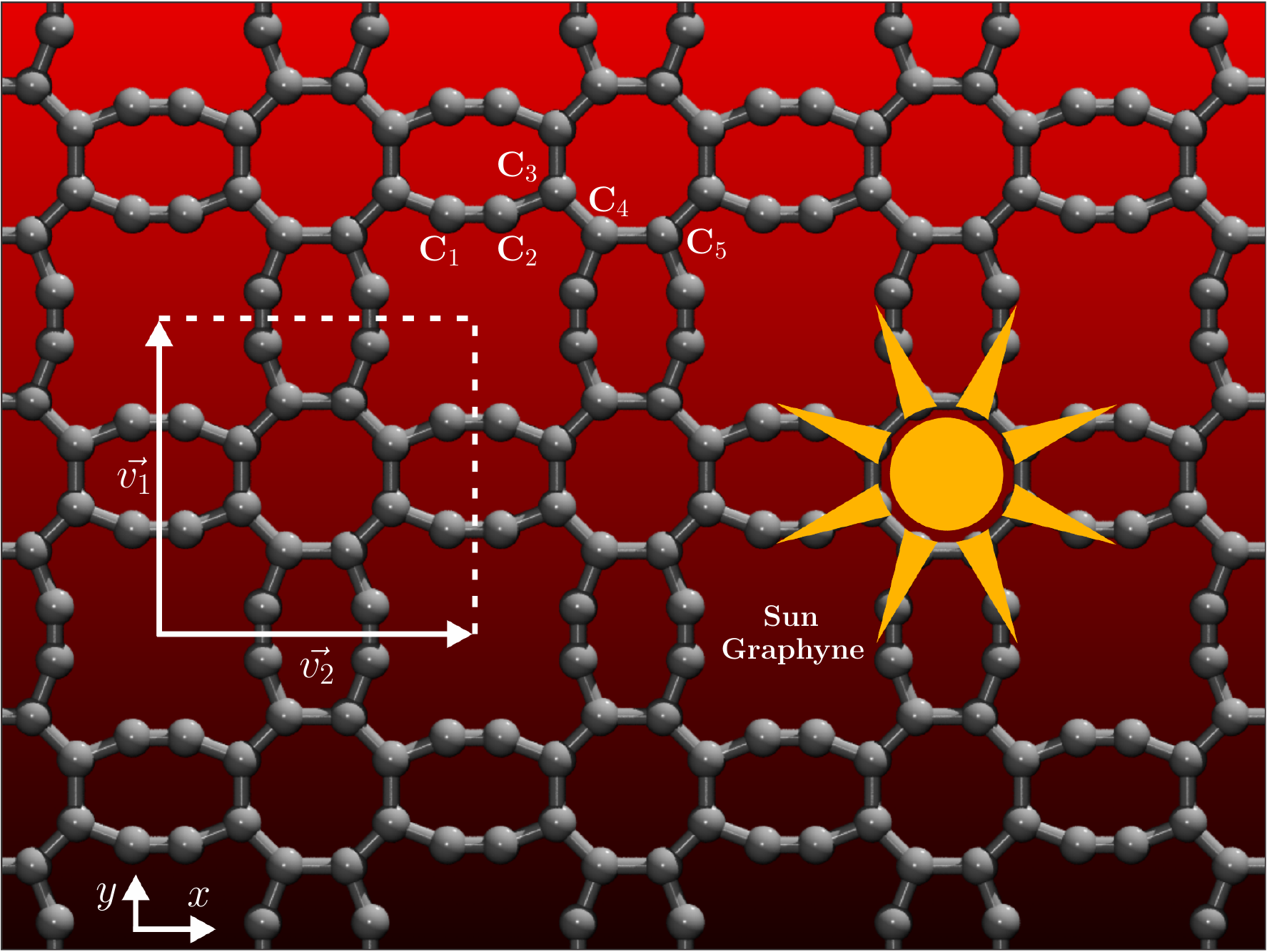}
\caption{Schematic representation of the S-GY structure. The symmetric unit cell is highlighted by the square.The unit cell vectors are $|\vec{v}_1|=|\vec{v}_2| = 7.36$ \r{A}.}
\label{fig1}
\end{figure*}

We also performed classical reactive MD simulations to investigate the S-GY fracture patterns, dynamics, and thermal stability. The fully atomistic MD simulations were performed using the Large-scale Atomic/Molecular Massively Parallel Simulator (LAMMPS) code \cite{thompson2022lammps,plimpton_JCP}. A reactive force field is required since fracture involves bond break and formation. For this reason, we used the Adaptive Intermolecular Reactive Empirical Bond Order potential (AIREBO) \cite{stuart2000reactive}.  

Newton's equations of motion were integrated using the Velocity-Verlet Algorithm. The integration time step was 0.1 fs, and an NPT ensemble was considered using the Nos\'e-Hoover thermostat \cite{hoover1985canonical}. The simulations were performed from room temperature up to 7000 K and null pressure. The S-GY lattice was initially equilibrated at 200 ps to eliminate residual stress. 

The systems considered here have an area of approximately 104 nm$^2$, consisting of a $14\times14\times1$ (3136 atoms) unit cell (which contains 16 carbon atoms), indicated in Figure \ref{fig1}, and periodic boundary conditions. Along the $z$-direction, a 100 \r{A} lattice constant (vacuum buffer zone) was used to avoid the interaction between the S-GY and its images.

For the strain simulations, only the $x$-direction was considered due to the symmetry of the S-GY unit cell (see Figure \ref{fig1}). The strain simulations were carried out by applying external stress by increasing the simulation box size with a constant uniaxial strain rate of $10^{-6}$ fs$^{-1}$ at room temperature. We can obtain the stress-strain curves and elastic properties from this simulation protocol. From representative MD snapshots, we can analyze the fracture patterns and dynamics.

The thermal stability of the S-GY was investigated using a heating ramp protocol. The temperature was increased from room temperature up to 7000K using a constant rate of 2 K/ps and an NVT ensemble during 1 ns. From the obtained results, we estimated the S-GY melting point. The MD snapshots and trajectories were obtained using the visualization and analysis software VMD \cite{HUMPHREY199633}.

\section{Results}

\subsection{Stability and Structural Properties}

The schematic representation of the optimized S-GY lattice is shown in Figure \ref{fig1}. Its symmetric unit cell is highlighted in the square. The unit cell vectors are $|\vec{v_1}|=|\vec{v_2}| = 7.36$ \r{A}. S-GY has an all-carbon structure periodically arranged with two different eight-atom rings. The bond length values are $\overline{\mathrm{C}_1\mathrm{C}_2}=1.24$ \r{A} and $\overline{\mathrm{C}_2\mathrm{C}_3}=1.42$ \r{A} for one ring, and $\overline{\mathrm{C}_3\mathrm{C}_4}=1.42$ \r{A} and $\overline{\mathrm{C}_4\mathrm{C}_5}=1.48$  \r{A} for the other ring, as shown in Figure \ref{fig1}. These values are typically found in similar 2D carbon allotropes \cite{Heimann1997,Enyashin2011}. It is worth mentioning that S-GY formation energy is -8.57 eV/atom, similar to graphene (-8.8 eV/atom) \cite{eform_wang,Tromer2017,nhg_boron}.

\begin{figure*}[!htb]
\centering
\includegraphics[width=0.7\linewidth]{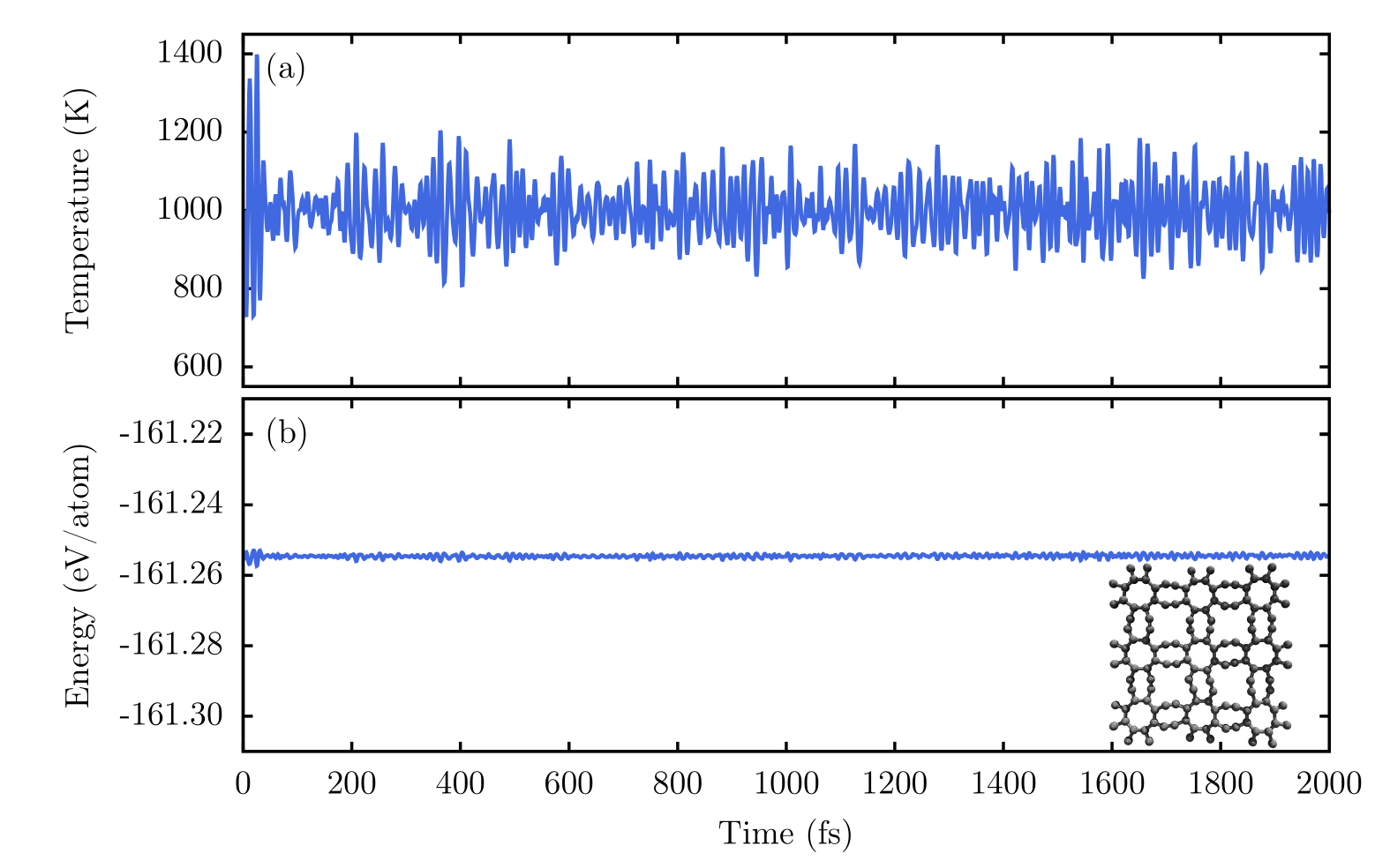}
\caption{Representative AIMD snapshots corresponding to the initial and final moments of the simulation. Here we considered a total time of 2 ps and a temperature of 1000 K.}
\label{fig2}
\end{figure*}

To test the S-GY structural stability, we performed AIMD simulations considering a total time of 2 ps at a temperature of 1000 K. In Figure \ref{fig2}, we show the representative AIMD snapshot corresponding to the final moment of the simulation. This figure shows that the S-GY structure remains practically unchanged at high-temperature regimes, with only small perturbations (some of the rings are slightly tilted) due to thermal fluctuations. The calculated formation energy value and the AIMD results indicate that S-GY is structurally stable. The recent advances in the graphyne syntheses \cite{Desyatkin2022,hu2022synthesis,barua2022novel} suggest that S-GY is possibly synthesizable.

\subsection{Electronic Properties}

\begin{figure}[!htb]
\centering
\includegraphics[width=\linewidth]{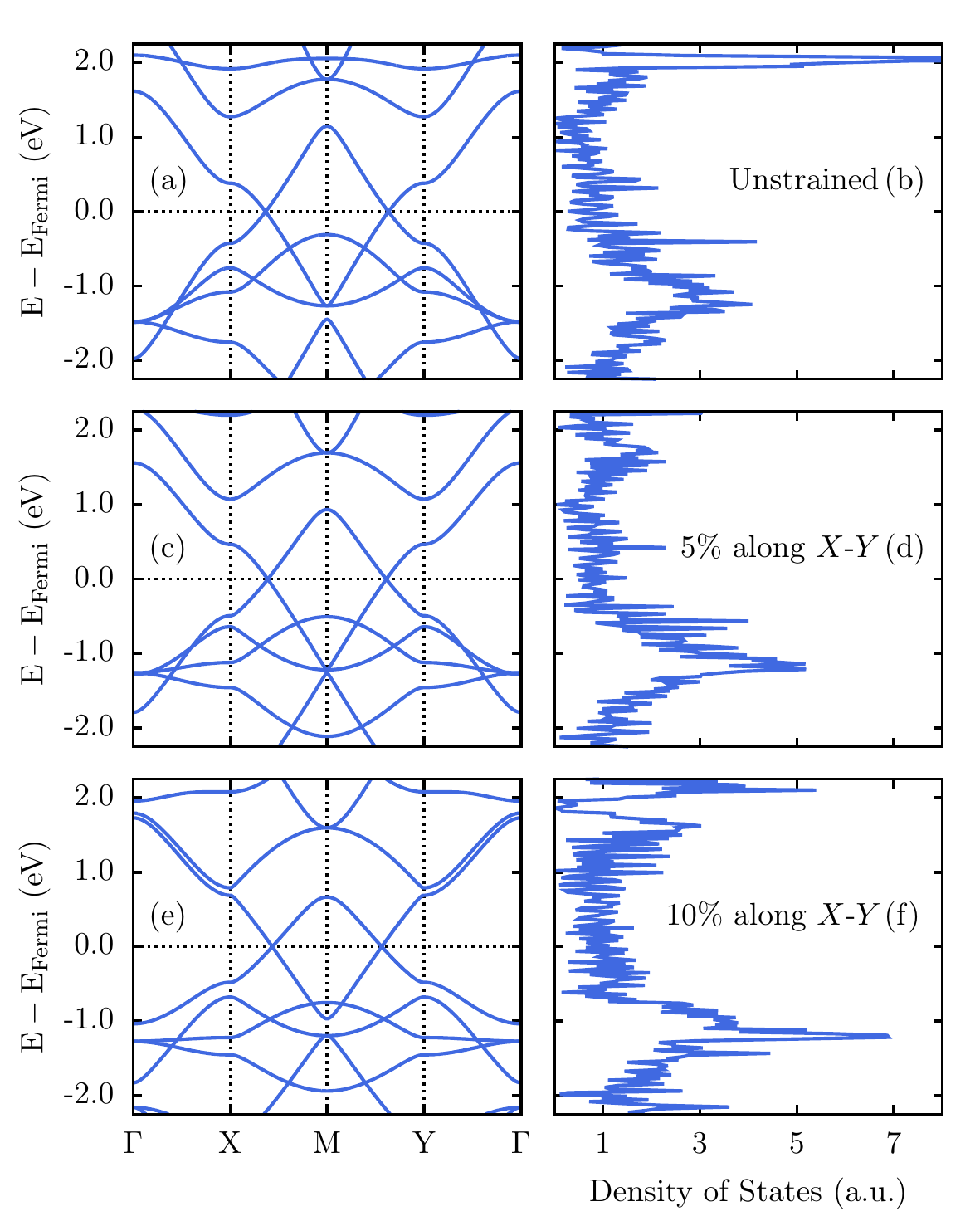}
\caption{S-GY electronic band structure for the unstrained (a-b), and with (c-d) 5\% and (e-f) 10\% of applied biaxial-strain.}
\label{fig3}
\end{figure}

In Figure \ref{fig3}, we present the S-GY electronic band structure for the unstrained (\ref{fig3}(a-b)) and with biaxial-strain (for \ref{fig3}(c-d) 5\% and \ref{fig3}(e-f) 10\% of strain) cases. Their corresponding density of states is also presented. As a general trend, we can see two Dirac points at the middle $X\rightarrow M$ and $M\rightarrow Y$ integration paths, with a narrow gap of about 5 meV. This feature indicates that the S-GY electrons would behave as massless Dirac fermions, similar to the graphene case. We can also see that S-GY is symmetric from $M\rightarrow \Gamma$ or $M \leftarrow \Gamma$, presenting isotropic transport channels.

We can see that the S-GY electronic band structure presented in Figures \ref{fig3}(a-b) remains practically unchanged even for moderate strain regimes. There is no symmetry inversion as in other 2D carbon-based materials with Dirac's cones. Figures \ref{fig3}(c-d) and \ref{fig3}(e-f) present the S-GY band structure for 5\% and 10\% for the biaxial strain, respectively. Both band configurations resemble the unstrained case. In other graphyne materials \cite{Wang2014_C}, the band structure can be tuned under the applied stress. In these cases, the two Dirac cones merge into only one cone for a biaxial strain of 6\% \cite{Wang2014_C}. Moreover, a bandgap was observed for a strain value of 8\% \cite{Wang2014_C}, contrasting with the present case, where a Dirac cone remains stable even for larger strain values.  

\begin{figure}[!htb]
\centering
\includegraphics[width=0.8\linewidth]{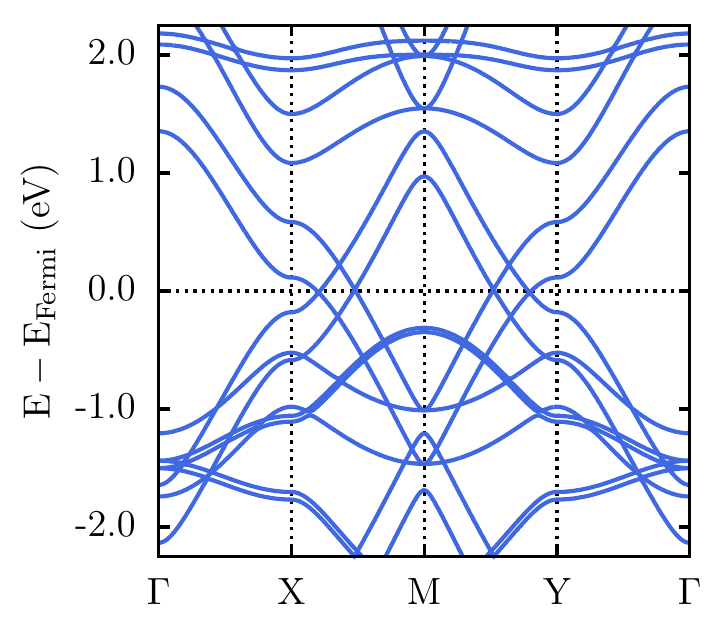}
\caption{Electronic band structure configuration for the S-GY bilayer.}
\label{fig4}
\end{figure}

To further investigate the band configuration of S-GY, we also analyzed the effect of the bilayer interactions on the electronic properties, as shown in Figure \ref{fig4}. As can be seen from this figure, the leading electronic features are similar to the monolayer case but with two extra Dirac cones. The new cones denote the contribution of an additional layer to the electronic properties. 

\subsection{Optical Properties}

\begin{figure}[!htb]
\centering
\includegraphics[width=\linewidth]{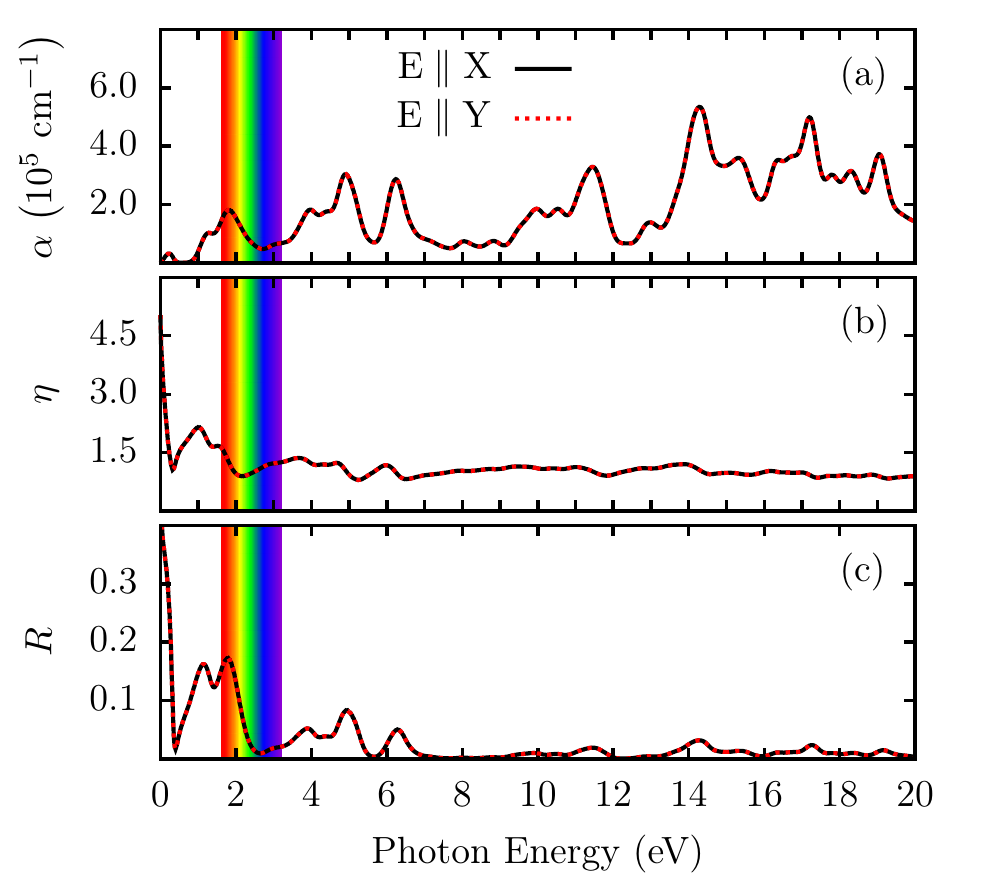}
\caption{(a) Absorption coefficient, (b) refractive index, and (c) reflectivity as a function of photon energy for the S-GY monolayer. $E||X$ and $E||Y$ denote the polarization direction for the externally applied electric field.}
\label{fig5}
\end{figure}

In Figure \ref{fig5}, we present the optical coefficients as a function of photon energy from 0 to 20 eV with an external electrical field polarized applied along the x and y directions. The absorption (Figure \ref{fig5}(a)) starts close to 0 eV ($\pi$ to $\pi^{*}$ transitions, inferred from the density of states). This trend is expected since S-GY is a semi-metallic material with an electronic bandgap of 5 meV, as mentioned above. We observe several peaks from infrared to ultraviolet regions. The maximum absorption intensity is about $3\times 10^{5}$cm$^{-1}$ for photon energies up to 12 eV and increases to $5\times 10^{5}$cm$^{-1}$ for higher photon energies. There is one peak within the visible region at 1.9 eV. We can also observe that S-GY has isotropic optical properties along the x and y directions, as expected from its topology.

\begin{figure}[!htb]
\centering
\includegraphics[width=\linewidth]{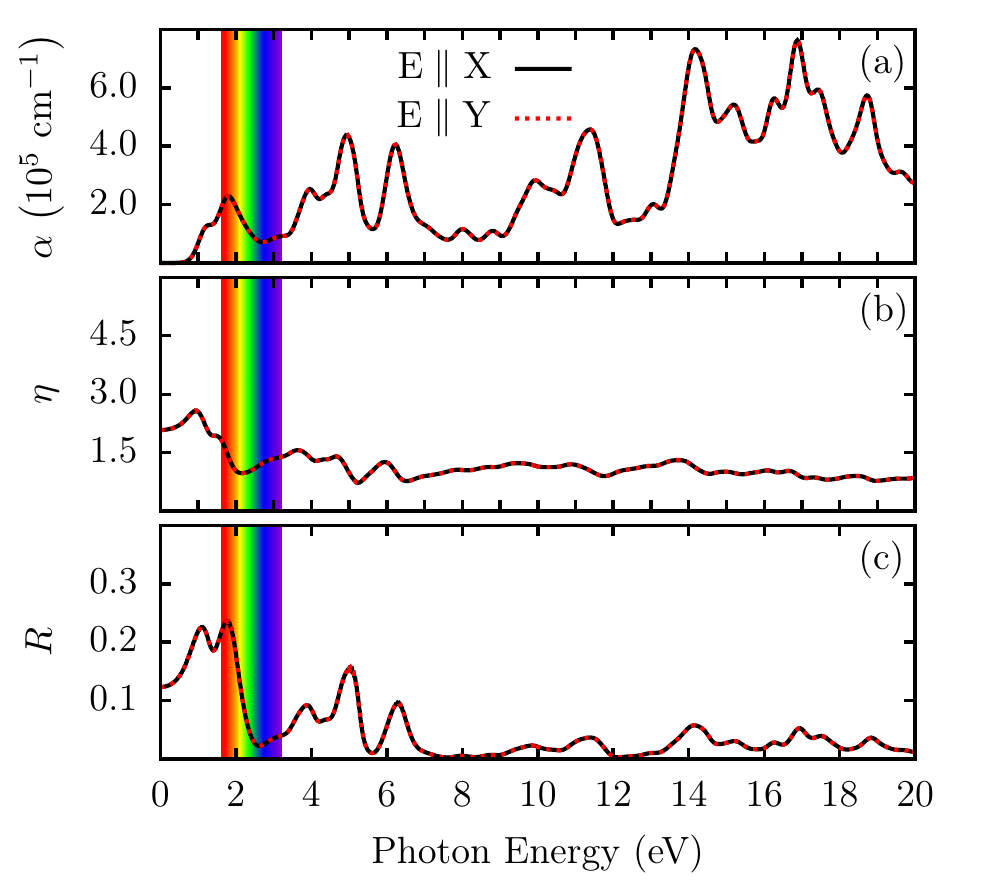}
\caption{(a) Absorption coefficient, (b) refractive index, and (c) reflectivity as a function of photon energy for the S-GY bilayer. $E||X$ and $E||Y$ denote the polarization direction for the externally applied electric field.}
\label{fig6}
\end{figure}
 
\begin{figure}[!htb]
\centering
\includegraphics[width=\linewidth]{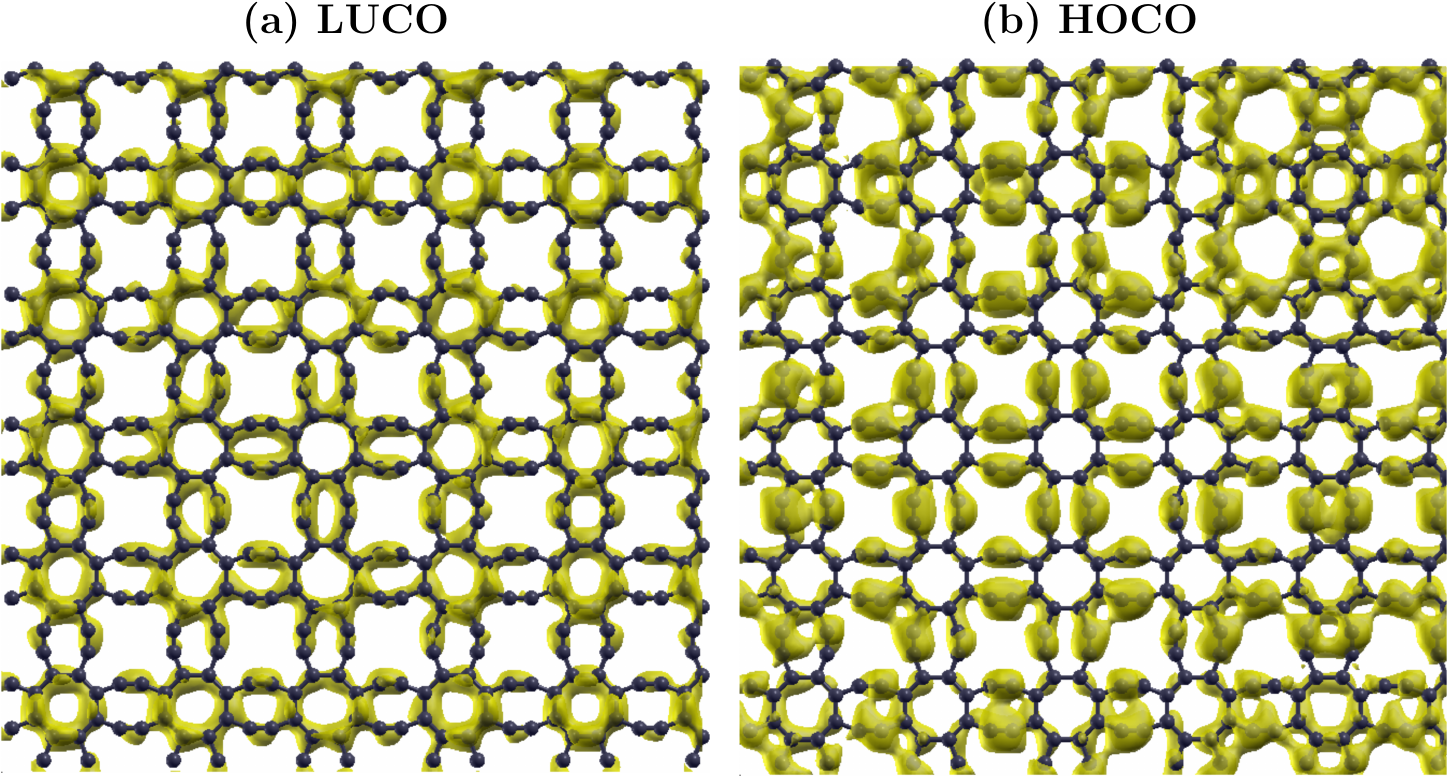}
\caption{Spatial distribution for the (a) LUCO and (b) HOCO.}
\label{fig7}
\end{figure} 

The refractive index is shown in Figure \ref{fig5}(b). Except for its value at null photon energy, the maximum intensity corresponds to the peak at 1 eV. $\eta$ slightly decreased up to 2 eV, remaining practically constant for all remaining spectra. A similar trend is observed for the S-GY reflectivity, as shown in Figure \ref{fig5}(c). Note that the maximum intensity for $R$ occurs between 1 and 2 eV. The $R$ activity is limited to the infrared region and decreases to values near zero for photon energies higher than 3 eV. These results suggest that the incident light on S-GY is almost entirely absorbed, i.e., it is a transparent material.

We also performed the same optical analysis for the bilayer case. As expected, we do not observe substantial differences in the optical activity between the monolayer and bilayer cases, as illustrated in Figure \ref{fig6}. The bilayer case has higher values of light absorption, about $8\times 10^{5}$cm$^{-1}$ (see Figure \ref{fig6}(a)). $R$ and $\eta$ behave similarly between the two systems (see Figures \ref{fig6}(b) and \ref{fig6}(c), respectively), except at null photon energy.
 
We analyzed the spatial patterns of the frontier crystalline orbitals. The results for the lowest unoccupied crystalline orbital (LUCO) and the highest occupied crystalline orbital (HOCO) are presented in Figure \ref{fig7}(a) and \ref{fig7}(b), respectively. As a general trend, both crystalline orbitals are spread over the lattice, consistent with electronic delocalization and the small bandgap value.
 
\subsection{Mechanical Properties}

\begin{figure}[!htb]
\centering
\includegraphics[width=0.8\linewidth]{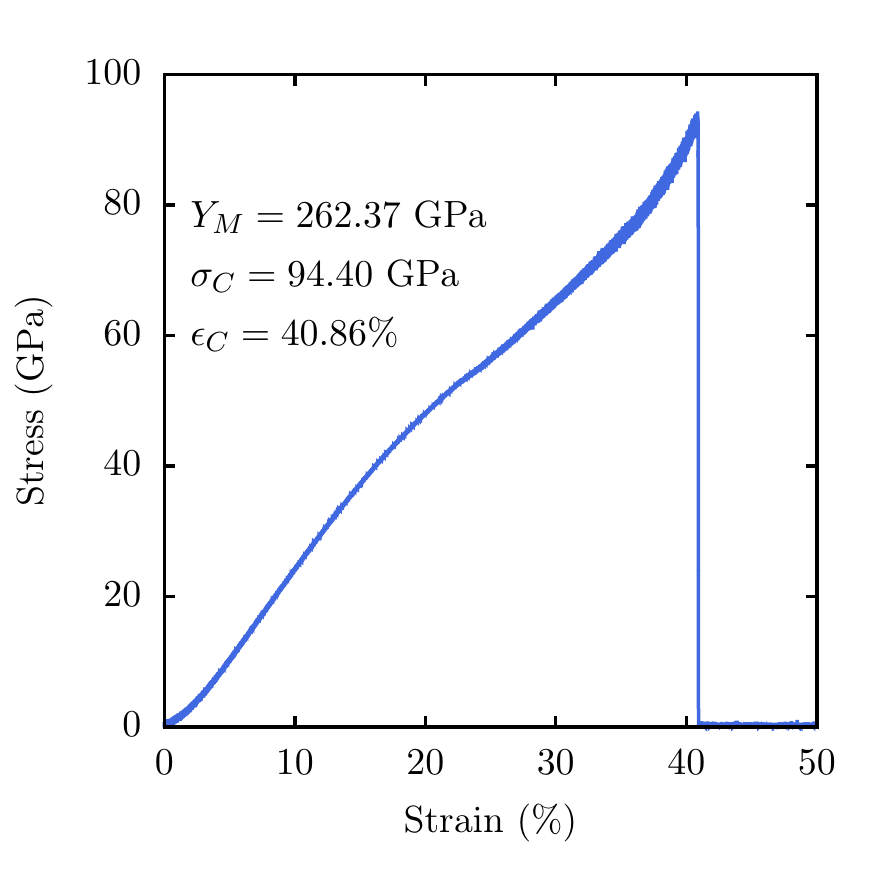}
\caption{Stress-strain curve for S-GY as a function of the uniaxial applied strain along the x-direction.}
\label{fig8}
\end{figure}

Figure \ref{fig8} presents the stress-strain curve for the uniaxial (x-direction) tensile loading. As the S-GY lattice is symmetric along the x and y directions, its mechanical properties are isotropic. 

S-GY exhibits a quasi-linear elastic region when subjected to stain values up to 40\%. It undergoes an abrupt transition involving a fractured configuration (null stress) after a critical strain ($\epsilon_C$) of 40.86\%, as depicted in Figure \ref{fig8}. The ultimate stress value ($\sigma_C$) is 94.40 GPa, considerably smaller than the corresponding graphene one (about 228.72 GPa \cite{felix2020mechanical}). The ultimate stress is the corresponding tensile stress for the critical strain. 

We considered 2\% of stress for calculating Young's modulus ($Y_M$), which is 262.37 GPa (or 87.6 N/m if we consider the structure's thickness as 3.34 \AA). This value is significantly smaller than that of graphene, and other similar 2D carbon-based structures \cite{felix2020mechanical,pereira2022mechanical}. In particular, comparing the value of the S-GY's Young's modulus with that of other known graphyne structures, it is about half, the same and twice that of $\gamma$-graphyne, $\beta$-graphyne, and $\alpha$-graphyne, respectively \cite{alexandre2022ctrends}. The high S-GY porosity can explain these differences.

\begin{figure*}[!htb]
\centering
\includegraphics[width=0.7\linewidth]{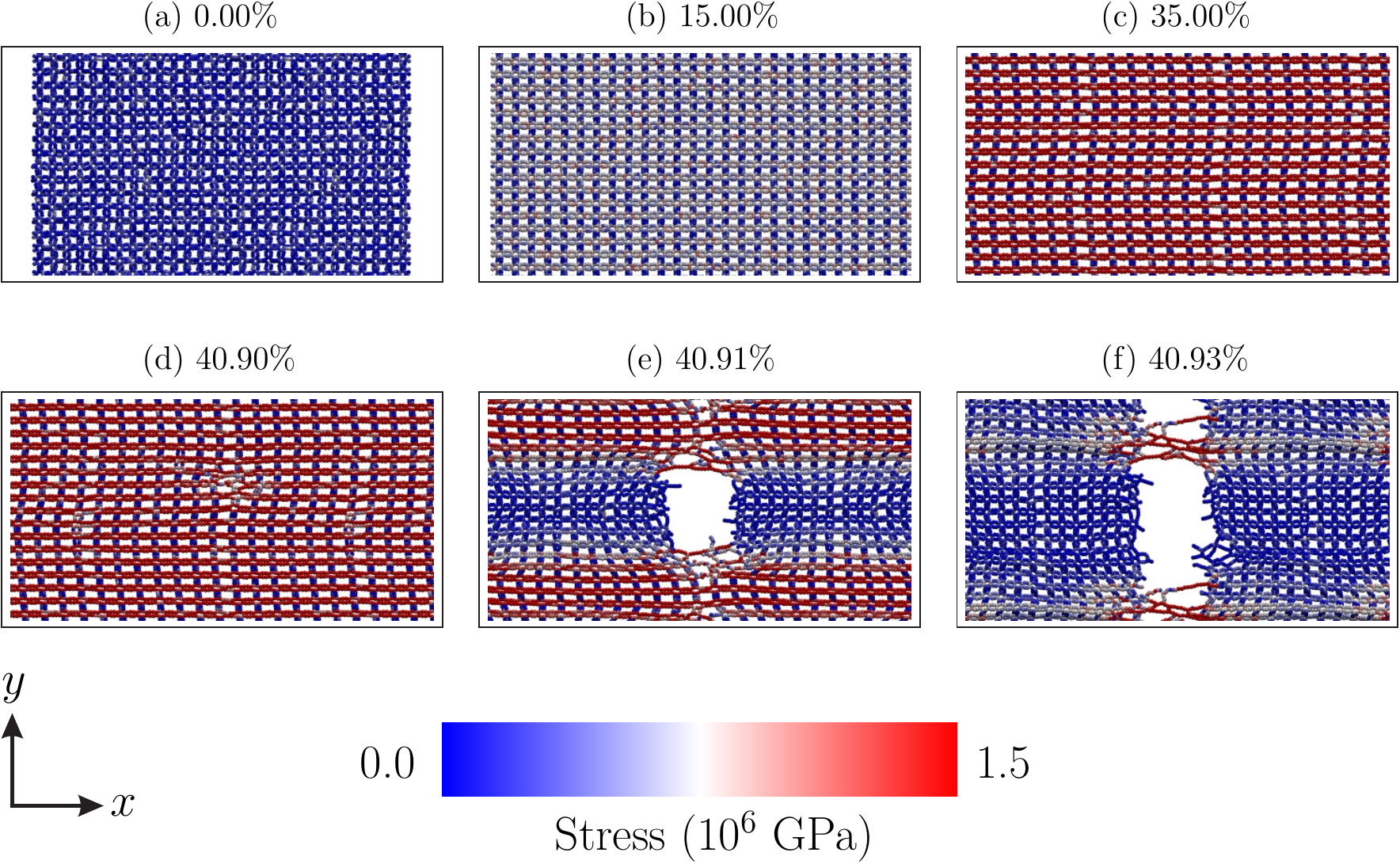}
\caption{Representative MD snapshots for the S-GY subjected to a strain applied along the x-direction.}
\label{fig9}
\end{figure*}

The S-GY fracture patterns and dynamics were also investigated here. In Figure \ref{fig9}(a-f), we show some representative MD snapshots highlighting critical moments of the stress dynamics. The colour scheme denotes the values for the von Mises stress per atom along the structure. These values provide helpful information on the fracture dynamics \cite{felix2020mechanical}. 

Figures \ref{fig9}(a) at 0.0\% of strain, \ref{fig9}(b) at 15.0\% of strain, and \ref{fig9}(c) show that S-GY can preserve its structural integrity up to 35\% of strain. The first bond break occurs at 40.90\% of strain, as shown in Figure \ref{fig9}(d). After this critical value, the lattice undergoes an abrupt brittle-like fracture with fast and linear crack propagation along the perpendicular direction of the stretch at 40.91\% of strain (see Figure \ref{fig9}(e)). This process separates the S-GY lattice into two parts connected by linear atomic chains (LACs) at 40.93\% of strain, as illustrated in Figure \ref{fig9}(f). The fracture starts from the acetylene bonds. The whole process can be better understood from video01 in the Supplementary Material.

\subsection{Melting Point}

\begin{figure}[!htb]
\centering
\includegraphics[width=0.8\linewidth]{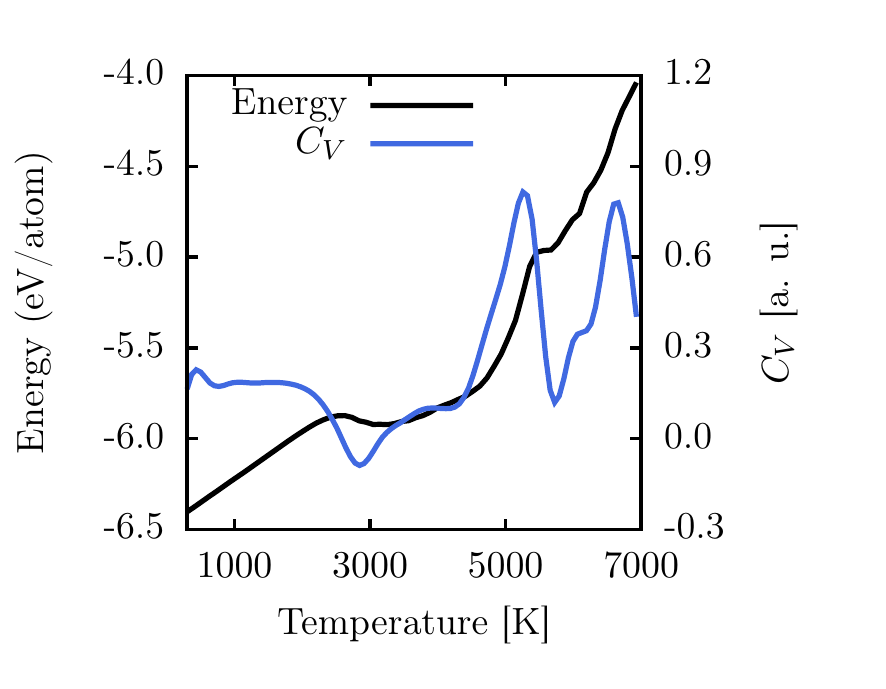}
\caption{Total energy (black) and heat capacity ($C_V$, blue) values as a function of temperature for the S-GY monolayer.}
\label{fig10}
\end{figure}

The melting point analysis was performed using the heating ramp protocol. In Figure \ref{fig10}, we present the total energy (black) and heat capacity ($C_V$, in blue) values as a function of temperature. The total energy increases quasi-linearly with temperature values between 300K-2100K, quasi-parabolic between 2100K-5500K, and quasi-linearly again for temperatures ranging within the interval 5500K-7000K. 

The most pronounced peak in the CV curve denotes a melting point of about 2800 K (see Figure \ref{fig10}). In the first stage of the melting process (300K-2100K), the S-GY lattice retains its integrity. At 2800K, the thermal vibrations lead to morphology changes, and the melting process occurs at the second heating stage (2100K-5500K). S-GY melting point is smaller than those for the monolayer graphene (4095K) \cite{los2015melting}, monolayer amorphous carbon (3626K) \cite{felix2020mechanical}, and biphenylene network (4024K) \cite{pereira2022mechanical}. The last stage of the heating process (between 5500K-7000K), with a significant change in the slope for the total energy curve, is associated with the complete structural destruction of the structure and its conversion to a gas phase. 

\begin{figure*}[!htb]
\centering
\includegraphics[width=0.7\linewidth]{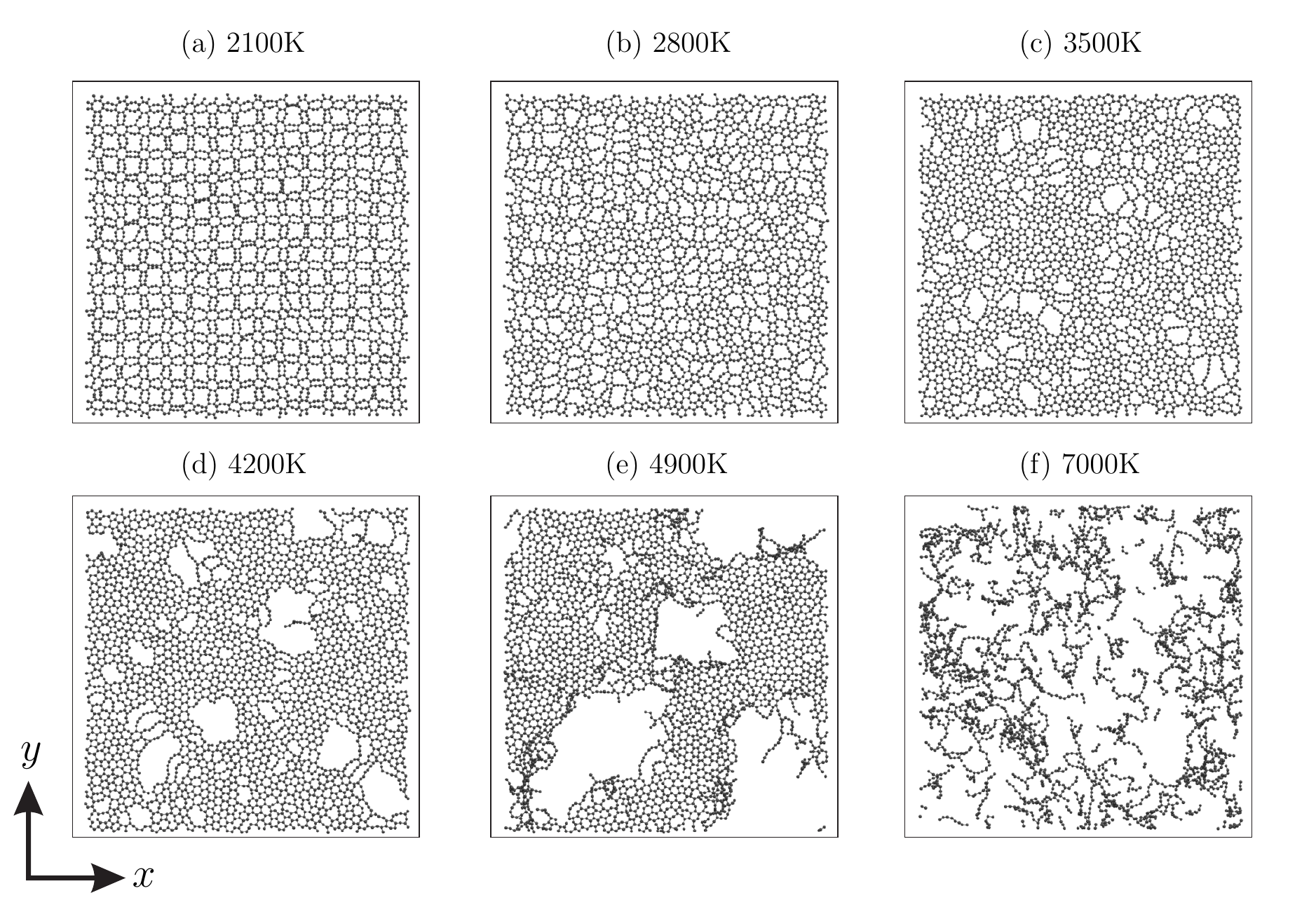}
\caption{Representative MD snapshots for the heating ramp simulations (melting process) for S-GY at (a) 2100K, (b) 2800K, (c) 4200K, (d) 4900K, (e) 4000K, and (f) 7000K.}
\label{fig11}
\end{figure*}

Finally, in Figure \ref{fig11}, we show representative MD snapshots for the heating ramp simulation of S-GY. The temperatures vary from 2100K up to 7000K. In Figure \ref{fig11}(a), we can see that the thermal vibrations lead to  changes in the lattice morphology with several C--C bond breakings and reconstructions. However, the overall structural configuration is still similar to the S-GY topology. The complete amorphization of the lattice occurs at 2800K, as shown in Figure \ref{fig11}(b). 

For temperatures between 3500K-5000K, we observe the formation of graphene-like domains, i.e., lattice fragments composed of six-membered rings, as illustrated in Figures \ref{fig11}(c-e). The complete atomization of the lattice occurs for temperatures higher than 5000K (see Figure \ref{fig11}(f)). The whole process can be better understood from video02 of the supplementary material.

\section{Conclusions}

We used DFT and reactive fully atomistic MD simulations to propose a new 2D carbon allotrope named Sun-Graphyne. This material has an all-carbon structure periodically arranged by two eight-atom carbon rings. 

We investigated the S-GY thermal and structural stability. AIMD simulations confirmed its structural and thermal stabilities. In these simulations, the S-GY retains its structural morphology up to 1000K. Its DFT formation energy is -8.57 eV/atom, similar to the graphene one (-8.8 eV/atom). These results suggest that S-GY is structurally stable.

Electronic structure calculations revealed that S-GY is a semi-metal material with a narrow gap of about 5 meV. Interestingly, its electronic band structure presents two Dirac cones and isotropic transport channels. The Dirac cones indicate that the electrons behave as massless Dirac fermions, similar to graphene. 

The S-GY electronic band structure remains unchanged even for moderate strain regimes, which, as far as we know, is unique for 2D carbon allotropes. Also, there is no symmetry inversion under strain as in similar 2D carbon-based materials with Dirac's cones.

S-GY has isotropic optical properties along the plane directions. Its reflectivity is limited to the infrared region and decreases to values near zero for photon energies higher than 3 eV. The incident light on its surface is almost entirely absorbed, i.e., this material is transparent.

Regarding the mechanical properties, S-GY possesses a quasi-linear elastic region when subjected to stain values up to 40\%. It abruptly transitions to a fractured state (null stress) after a critical strain of 40.86\%. Its ultimate stress value and Young's modulus are 94.40 GPa and 262.37 GPa, respectively. These values are considerably smaller than the graphene ones \cite{felix2020mechanical} and comparable to those of other known graphyne structures \cite{alexandre2022ctrends}. Moreover, these values are much lower than similar 2D carbon-based structures, which can be attributed to the S-GY porosity.

The S-GY melting point (2800K) is smaller than those for the monolayer graphene (4095K) \cite{los2015melting}, monolayer amorphous carbon (3626K) \cite{felix2020mechanical}, and biphenylene network (4024K) \cite{pereira2022mechanical}. For temperatures between 3500K-5000K, the S-GY melting process tends to form graphene-like domains. 

Considering recent advances in graphyne synthesis, S-GY exhibits some unique properties and is possibly synthesizable. We hope the present study can stimulate further studies for this remarkable new structure.

\section*{Acknowledgements}

This work was financed by the Coordenaç\~ao de Aperfeiçoamento de Pessoal de Nível Superior (CAPES) - Finance Code 001 and grant 88887.691997/2022-00, Conselho Nacional de Desenvolvimento Científico e Tecnol\'ogico (CNPq), FAP-DF, and FAPESP. We thank the Center for Computing in Engineering and Sciences at Unicamp for financial support through the FAPESP/CEPID Grants \#2013/08293-7 and \#2018/11352-7. L.A.R.J acknowledges the financial support from FAP-DF grants $00193-00000857/2021-14$, $00193-00000853/2021-28$, $00193-00000811/2021-97$, and, $00193.00001808/2022-71$ CNPq grants $302922/2021-0$ and $350176/2022-1$. L.A.R.J. gratefully acknowledges the support from ABIN grant 08/2019 and Fundaç\~ao de Apoio \ 'a Pesquisa (FUNAPE), Edital 02/2022 - Formul\'ario de Inscriç\~ao N.4. L.R.S. acknowledges the National Institute of Science and Technology of Complex Systems (Brazil). L.A.R.J. acknowledges N\'ucleo de Computaç\~ao de Alto Desempenho (NACAD) and for providing the computational facilities. This work used resources of the Centro Nacional de Processamento de Alto Desempenho em S\~ao Paulo (CENAPAD-SP). The authors acknowledge the National Laboratory for Scientific Computing (LNCC/MCTI, Brazil) for providing HPC resources of the SDumont supercomputer, contributing to the research results reported within this paper. URL: http://sdumont.lncc.br. 

\bibliography{references}

\end{document}